\def\pp{{\mathchoice
          {%displaystyle
              \kern 1pt%
              \raise 1pt
              \vbox{\hrule width5pt height0.4pt depth0pt
                    \kern -2pt
                    \hbox{\kern 2.3pt
                          \vrule width0.4pt height6pt depth0pt
                          }
                    \kern -2pt
                    \hrule width5pt height0.4pt depth0pt}%
                    \kern 1pt
           }
            {%textstyle
              \kern 1pt%
              \raise 1pt
              \vbox{\hrule width4.3pt height0.4pt depth0pt
                    \kern -1.8pt
                    \hbox{\kern 1.95pt
                          \vrule width0.4pt height5.4pt depth0pt
                          }
                    \kern -1.8pt
                    \hrule width4.3pt height0.4pt depth0pt}%
                    \kern 1pt
            }
            {%scriptstyle
              \kern 0.5pt%
              \raise 1pt
              \vbox{\hrule width4.0pt height0.3pt depth0pt
                    \kern -1.9pt  %[e]=0.15pt
                    \hbox{\kern 1.85pt
                          \vrule width0.3pt height5.7pt depth0pt
                          }
                    \kern -1.9pt
                    \hrule width4.0pt height0.3pt depth0pt}%
                    \kern 0.5pt
            }
            {%scriptscriptstyle
              \kern 0.5pt%
              \raise 1pt
              \vbox{\hrule width3.6pt height0.3pt depth0pt
                    \kern -1.5pt
                    \hbox{\kern 1.65pt
                          \vrule width0.3pt height4.5pt depth0pt
                          }
                    \kern -1.5pt
                    \hrule width3.6pt height0.3pt depth0pt}%
                    \kern 0.5pt%}
            }
        }}

\def\mm{{\mathchoice
                       {%displaystyle
                             \kern 1pt
               \raise 1pt    \vbox{\hrule width5pt height0.4pt depth0pt
                                  \kern 2pt
                                  \hrule width5pt height0.4pt depth0pt}
                             \kern 1pt}
                       {%textstyle
                            \kern 1pt
               \raise 1pt \vbox{\hrule width4.3pt height0.4pt depth0pt
                                  \kern 1.8pt
                                  \hrule width4.3pt height0.4pt depth0pt}
                             \kern 1pt}
                       {%scriptstyle
                            \kern 0.5pt
               \raise 1pt
                            \vbox{\hrule width4.0pt height0.3pt depth0pt
                                  \kern 1.9pt
                                  \hrule width4.0pt height0.3pt depth0pt}
                            \kern 1pt}
                       {%scriptscriptstyle
                           \kern 0.5pt
             \raise 1pt  \vbox{\hrule width3.6pt height0.3pt depth0pt
                                  \kern 1.5pt
                                  \hrule width3.6pt height0.3pt depth0pt}
                           \kern 0.5pt}
                       }}

\documentstyle[12pt]{article}

\skewchar\fivmi='177
\skewchar\sixmi='177
\skewchar\sevmi='177
\skewchar\egtmi='177
\skewchar\ninmi='177
\skewchar\tenmi='177
\skewchar\elvmi='177
\skewchar\twlmi='177
\skewchar\frtnmi='177
\skewchar\svtnmi='177
\skewchar\twtymi='177
\def\@magscale#1{ scaled \magstep #1}
\skewchar\fivsy='60
\skewchar\sixsy='60
\skewchar\sevsy='60
\skewchar\egtsy='60
\skewchar\ninsy='60
\skewchar\tensy='60
\skewchar\elvsy='60
\skewchar\twlsy='60
\skewchar\frtnsy='60
\skewchar\svtnsy='60
\skewchar\twtysy='60

\catcode`@=11
\def\un#1{\relax\ifmmode\@@underline#1\else
        $\@@underline{\hbox{#1}}$\relax\fi}
\catcode`@=12

\def\a{\alpha}
\def\b{\beta}

\def\d{\delta}
\def\e{\epsilon}

\def\g{\gamma}

\def\k{\kappa}

\def\m{\mu}

\def\o{\omega}
\def\p{\pi}
\def\q{\theta}

\def\s{\sigma}
\def\t{\tau}

\def\z{\zeta}
\def\D{\Delta}

\def\G{\Gamma}

\def\L{\Lambda}

\def\P{\Pi}
\def\Q{\Theta}
\def\S{\Sigma}

\def\dslash{\not{\hbox{\kern-2pt $\partial$}}}
\def\Dslash{\not{\hbox{\kern-4pt $D$}}}
\def\pslash{\not{\hbox{\kern-2.3pt $p$}}}
 \newtoks\slashfraction
 \slashfraction={.13}
 \def\slash#1{\setbox0\hbox{$ #1 $}
 \setbox0\hbox to \the\slashfraction\wd0{\hss \box0}/\box0 }
 
\font\ro=cmsy10                          % font with rope
\def\kcr{{\hbox{\ro \char'170}}}                % right-handed rope
\def\ktl{{\hbox{\ro \char'170}}}        % top end for left-handed rope
\def\ktr{{\hbox{\ro \char'170}}}        % " right
\def\kbl{{\hbox{\ro \char'170}}}        % " bottom left
\def\kbr{{\hbox{\ro \char'170}}}        % " right
\def\plpl{\raise-2pt\hbox{$\raise3pt\hbox{$_+$}\hskip-6.67pt\raise0.0pt
\hbox{$^+$}\hskip 0.01pt$}}
\def\mimi{\raise-2pt\hbox{$\raise3pt\hbox{$_-$}\hskip-6.67pt\raise0.0pt
\hbox{$^-$}\hskip 0.01pt$}} 

\def\bo{{\raise.15ex\hbox{\large$\Box$}}}               % D'Alembertian
\def\pa{\partial}                                       % curly d
\def\TH{{\raise.2ex\hbox{$\displaystyle \bigodot$}\mskip-4.7mu \llap H \;}}
\def\face{{\raise.2ex\hbox{$\displaystyle \bigodot$}\mskip-2.2mu \llap {$\ddot
        \smile$}}}                                      % happy face
\def\pp{{\mathchoice
          {%displaystyle
              \kern 1pt%
              \raise 1pt
              \vbox{\hrule width5pt height0.4pt depth0pt
                    \kern -2pt
                    \hbox{\kern 2.3pt
                          \vrule width0.4pt height6pt depth0pt
                          }
                    \kern -2pt
                    \hrule width5pt height0.4pt depth0pt}%
                    \kern 1pt
           }
            {%textstyle
              \kern 1pt%
              \raise 1pt
              \vbox{\hrule width4.3pt height0.4pt depth0pt
                    \kern -1.8pt
                    \hbox{\kern 1.95pt
                          \vrule width0.4pt height5.4pt depth0pt
                          }
                    \kern -1.8pt
                    \hrule width4.3pt height0.4pt depth0pt}%
                    \kern 1pt
            }
            {%scriptstyle
              \kern 0.5pt%
              \raise 1pt
              \vbox{\hrule width4.0pt height0.3pt depth0pt
                    \kern -1.9pt  %[e]=0.15pt
                    \hbox{\kern 1.85pt
                          \vrule width0.3pt height5.7pt depth0pt
                          }
                    \kern -1.9pt
                    \hrule width4.0pt height0.3pt depth0pt}%
                    \kern 0.5pt
            }
            {%scriptscriptstyle
              \kern 0.5pt%
              \raise 1pt
              \vbox{\hrule width3.6pt height0.3pt depth0pt
                    \kern -1.5pt
                    \hbox{\kern 1.65pt
                          \vrule width0.3pt height4.5pt depth0pt
                          }
                    \kern -1.5pt
                    \hrule width3.6pt height0.3pt depth0pt}%
                    \kern 0.5pt%}
            }
        }}

\def\sp#1{{}^{#1}}                              % superscript (unaligned)
\def\Tilde#1{\widetilde{#1}}                    % big tilde
\def\Hat#1{\widehat{#1}}                        % big hat
\def\Bar#1{\overline{#1}}                       % big bar
\def\leftrightarrowfill{$\mathsurround=0pt \mathord\leftarrow \mkern-6mu
        \cleaders\hbox{$\mkern-2mu \mathord- \mkern-2mu$}\hfill
        \mkern-6mu \mathord\rightarrow$}
\def\dvec#1{\vbox{\ialign{##\crcr
        \leftrightarrowfill\crcr\noalign{\kern-1pt\nointerlineskip}
        $\hfil\displaystyle{#1}\hfil$\crcr}}}           % <--> accent
\def\frac#1#2{{\textstyle{#1\over\vphantom2\smash{\raise.20ex
        \hbox{$\scriptstyle{#2}$}}}}}                   % fraction
\def\sfrac#1#2{{\vphantom1\smash{\lower.5ex\hbox{\small$#1$}}\over
        \vphantom1\smash{\raise.4ex\hbox{\small$#2$}}}} % alternate fraction
\def\bfrac#1#2{{\vphantom1\smash{\lower.5ex\hbox{$#1$}}\over
        \vphantom1\smash{\raise.3ex\hbox{$#2$}}}}       % "
\def\afrac#1#2{{\vphantom1\smash{\lower.5ex\hbox{$#1$}}\over#2}}    % "
\newskip\humongous \humongous=0pt plus 1000pt minus 1000pt
\def\caja{\mathsurround=0pt}
\def\eqalign#1{\,\vcenter{\openup2\jot \caja
        \ialign{\strut \hfil$\displaystyle{##}$&$
        \displaystyle{{}##}$\hfil\crcr#1\crcr}}\,}
\newif\ifdtup

\def\ref#1{$\sp{#1)}$}

\parskip=\medskipamount                 % space between paragraphs (LaTeX)
\thicklines                         % thick straight lines for pictures (LaTeX)

\def\oldheadpic{                                % old UM heading
        \setlength{\unitlength}{.4mm}
        \thinlines
        \par
        \begin{picture}(349,16)
        \put(325,16){\line(1,0){4}}
        \put(330,16){\line(1,0){4}}
        \put(340,16){\line(1,0){4}}
        \put(335,0){\line(1,0){4}}
        \put(340,0){\line(1,0){4}}
        \put(345,0){\line(1,0){4}}
        \put(329,0){\line(0,1){16}}
        \put(330,0){\line(0,1){16}}
        \put(339,0){\line(0,1){16}}
        \put(340,0){\line(0,1){16}}
        \put(344,0){\line(0,1){16}}
        \put(345,0){\line(0,1){16}}
        \put(329,16){\oval(8,32)[bl]}
        \put(330,16){\oval(8,32)[br]}
        \put(339,0){\oval(8,32)[tl]}
        \put(345,0){\oval(8,32)[tr]}
        \end{picture}
        \par
        \thicklines
        \vskip.2in}
\def\oldtitle#1#2#3#4{\oldheadpic\begin{center}\vglue.5in{\large\bf #1}\\[.6in]
        {#2}\\[.1in] {\it Department of Physics and Astronomy}\\
        {\it University of Maryland, College Park, MD 20742}\\[.6in]
        Physics Publication \#{#3}\\ {#4}\\[1.5in] {\bf ABSTRACT}\\[.1in]
        \end{center} \begin{quotation}}                 % old title stuff
\def\oldTitle#1#2#3#4#5#6#7{\oldheadpic\begin{center} \vglue .4in
        {\large\bf #1}\\[.4in]
        {#2}\\[.1in] {\it Department of Physics and Astronomy}\\
        {\it University of Maryland, College Park, MD 20742}\\[.1in]
        {#3}\\[.1in] {\it {#4}}\\ {\it {#5}}\\[.4in]
        Physics Publication \#{#6}\\ {#7}\\[.5in] {\bf ABSTRACT}\\[.1in]
        \end{center} \begin{quotation}}                 % " for 2 authors
\def\border{                                            % border
        \setlength{\unitlength}{1mm}
        \newcount\xco
        \newcount\yco
        \xco=-21
        \yco=12
        \begin{picture}(140,0)
        \put(\xco,\yco){$\ktl$}
        \advance\yco by-1
        {\loop
        \put(\xco,\yco){$\kcr$}
        \advance\yco by-2
        \ifnum\yco>-240
        \repeat
        \put(\xco,\yco){$\kbl$}}
        \xco=158
        \yco=12
        \put(\xco,\yco){$\ktr$}
        \advance\yco by-1
        {\loop
        \put(\xco,\yco){$\kcr$}
        \advance\yco by-2
        \ifnum\yco>-240
        \repeat
        \put(\xco,\yco){$\kbr$}}
        \put(-20,13){\tiny University of Maryland Elementary Particle
Physics University of Maryland Elementary Particle Physics University of
Maryland Elementary Particle Physics}
        \put(-20,-241.5){\tiny University of Maryland Elementary
Particle Physics University of Maryland Elementary Particle Physics
University of Maryland Elementary Particle Physics}
        \end{picture}
        \par\vskip-8mm}
\def\bordero{                                           % alternate border
        \setlength{\unitlength}{1mm}
        \newcount\xco
        \newcount\yco
        \xco=-31
        \yco=12
        \begin{picture}(140,0)
        \put(\xco,\yco){$\ktl$}
        \advance\yco by-1
        {\loop
        \put(\xco,\yco){$\kclr}
        \advance\yco by-2
        \ifnum\yco>-240
        \repeat
        \put(\xco,\yco){$\kbl$}}
        \xco=151
        \yco=12
        \put(\xco,\yco){$\ktr$}
        \advance\yco by-1
        {\loop
        \put(\xco,\yco){$\kcr$}
        \advance\yco by-2
        \ifnum\yco>-240
        \repeat
        \put(\xco,\yco){$\kbr$}}
        \put(-20,12){\ooo bacdefghidfghghdhededbihdgdfdfhhdheidhdhebaaahjhhdahba

hgdedge
   hgfdiehhgdigicba}
        \put(-20,-241.5){\ooo ababaighefdbfghgeahgdfgafagihdidihiidhiagfedhadbfd

ecdcdfa
   gdcbhaddhbgfchbgfdacfediacbabab}
        \end{picture}
        \par\vskip-8mm}
\def\headpic{                                           % UM heading
        \indent
        \setlength{\unitlength}{.4mm}
        \thinlines
        \par
        \begin{picture}(29,16)
        \put(165,16){\line(1,0){4}}
        \put(170,16){\line(1,0){4}}
        \put(180,16){\line(1,0){4}}
        \put(175,0){\line(1,0){4}}
        \put(180,0){\line(1,0){4}}
        \put(185,0){\line(1,0){4}}
        \put(169,0){\line(0,1){16}}
        \put(170,0){\line(0,1){16}}
        \put(179,0){\line(0,1){16}}
        \put(180,0){\line(0,1){16}}
        \put(184,0){\line(0,1){16}}
        \put(185,0){\line(0,1){16}}
        \put(169,16){\oval(8,32)[bl]}
        \put(170,16){\oval(8,32)[br]}
        \put(179,0){\oval(8,32)[tl]}
        \put(185,0){\oval(8,32)[tr]}
        \end{picture}
        \par\vskip-6.5mm
        \thicklines}
\def\title#1#2#3#4{\border\headpic {\hbox to\hsize{#4 \hfill UMDEPP #3}}\par
        \begin{center} \vglue .5in {\large\bf #1}\\[.6in]
        {#2}\\[.1in] {\it Department of Physics and Astronomy}\\
        {\it University of Maryland, College Park, MD 20742}\\[1.5in]
        {\bf ABSTRACT}\\[.1in] \end{center} \begin{quotation}}  % title stuff
\def\Title#1#2#3#4#5#6#7{\border\headpic
        {\hbox to\hsize{#7 \hfill UMDEPP #6}}\par
        \begin{center} \vglue .4in {\large\bf #1}\\[.4in]
        {#2}\\[.1in] {\it Department of Physics and Astronomy}\\
        {\it University of Maryland, College Park, MD 20742}\\[.1in]
        {#3}\\[.1in] {\it {#4}}\\ {\it {#5}}\\[.5in] {\bf ABSTRACT}\\[.1in]
        \end{center} \begin{quotation}}                 % " for 2 authors
\def\endtitle{\end{quotation}\newpage}                  % end title page

\def\ad{{\kern0.5pt
                   \alpha \kern-5.05pt \raise5.8pt\hbox{$\textstyle.$}\kern
0.5pt}}

\def\bd{{\kern0.5pt
                   \beta \kern-5.05pt \raise5.8pt\hbox{$\textstyle.$}\kern
0.5pt}}

\def\qd{{\kern0.5pt
                   q \kern-5.05pt \raise5.8pt\hbox{$\textstyle.$}\kern
0.5pt}}

\begin{document}

\def\gfrac#1#2{\frac {\scriptstyle{#1}}
        {\mbox{\raisebox{-.6ex}{$\scriptstyle{#2}$}}}}
\def\gg{{\hbox{\sc g}}}
\border\headpic {\hbox to\hsize{November 1995 \hfill {UMDEPP 96-46}}}
\par
\setlength{\oddsidemargin}{0.3in}
\setlength{\evensidemargin}{-0.3in}
\begin{center}
\vglue .08in
{\large\bf On Aspects and Implications of the New Covariant\\
4D, N = 1 Green-Schwarz $\s$-model Action\footnote {Supported in 
part by National Science Foundation Grant PHY-91-19746 
\newline ${~~~~~}$ and by NATO 
Grant CRG-93-0789}  }
\\[.72in]

S. James Gates, Jr.
\\[.02in]
{\it Department of Physics\\ 
University of Maryland\\ 
College Park, MD 20742-4111  USA}\\[.2in] 
{\bf {\tt gates@umdhep.umd.edu}}\\[2.7in]

{\bf ABSTRACT}\\[.002in]
\end{center}
\begin{quotation}
{Utilizing (2,0) superfields, we write a (supersymmetry$)^2$ action and
partially relate it to the new formulation of the Green-Schwarz action
given by Berkovits and Siegel.  Recent results derived from this new 
formulation are discussed within the context of some prior proposals 
in the literature.  Among these, we note that 4D, N = 1 $\b$FFC 
superspace geometry with a composite connection for ${\cal R}$-symmetry 
has now been confirmed as the {\underline {only}} presently known 
limit of 4D, N = 1 heterotic string theory that is derivable in a 
completely rigorous manner.}  

\endtitle
%%%%%%%%%%%%%%%%%%%%%%%%%%%%%%%%%%%%%%%%%%%%%%%%%%%%%%%%%%%%%%%%%%
\section{Introduction} 
%%%%%%%%%%%%%%%%%%%%%%%%%%%%%%%%%%%%%%%%%%%%%%%%%%%%%%%%%%%%%%%%%%%%

~~~~Near the very beginning of our study of two-dimensional locally
supersymmetric field theories that are relevant to superstrings \cite{A},
we suggested the radical notion that the ultimate formulation of superstring
actions must be a bizarre hybrid of the Green-Schwarz (GS) and 
Neveu-Schwarz-Ramond (NSR) actions. The reason for making this proposal
was in answer to a simple but deeply troubling question, ``Why do the
two seemingly different formulations (NSR vs.~GS) both exist?'' It seemed
to us that the simplest resolution to this puzzle was the idea that the
two different formulations were really just two different ways (possibly
even different gauge choices) of looking at a meta-string formulation that
contained elements of both.  Hence, our seemingly fanciful suggestion. Our
original proposal can be put in the form of the (SUSY$)^2$ principle:

${~~~~~}${\it {The ultimate covariant formulation of superstring theory must involve
\newline ${~~~~~}$ a set of variables that describe a map from a world-sheet
supermanifold \newline ${~~~~~}$ into a space-time supermanifold.}}

\noindent The standard GS formulation may be thought of as a map from a 2d
world-sheet into a supermanifold and the NSR formulation may be thought
of as a map from a super 2d manifold into a bosonic manifold.  So our
suggestion was just the next logical progression.  In a series of
paper \cite{B}, we have explored various ways in which the realization
of the (SUSY$)^2$ principle might lead to an improved version of
superparticle and superstring theory.  As well, numbers of other authors
\cite{C} have investigated variations on this often-rediscovered   
``crazy idea'' \cite{D}.

A more recent development program \cite{E}, has occurred that yields a
result that comes tantalizingly close to the successful realization of
the (SUSY$)^2$ principle.  This is most evident in a recent paper \cite{F}.
A manifest, classical (SUSY$)^2$ realization is {\underline {not}} present.
However, at the quantum level, this model does apparently provide a
realization.  In this light, it is clearly an important object for
our study. In the present work, we will discuss aspects of this new $\s$-model
and show how it can be embedded into a classical (SUSY$)^2$ model.  We
will also see how this new model settles a number of issues that were
raised in previous investigations of 4D, N = 1 Green-Schwarz non-linear
$\s$-models \cite{G}. Perhaps the most important along these lines
is to note that \cite{F} provides an independent and rigorous 
derivation of the 4D, N = 1 supergravity limit. Namely, it is found that 
the 4D, N = 1 supergravity theory emerging from the heterotic string is
the ``old minimal'' supergravity theory ``entangled'' with
a tensor multiplet that acts as the composite connection for ${\cal 
R}$-symmetry exactly as described in reference \cite{G}.

%%%%%%%%%%%%%%%%%%%%%%%%%%%%%%%%%%%%%%%%%%%%%%%%%%%%%%%%%%%%%%%%%%%%
\section{Review of Local (2,0) Superspace Supergravity \newline
Geometry}
%%%%%%%%%%%%%%%%%%%%%%%%%%%%%%%%%%%%%%%%%%%%%%%%%%%%%%%%%%%%%%%%%%%

~~~~Some time ago, the geometry of (2,0) supergravity was developed
\cite{H}.  At that time, however, the formulation did not take advantage
of the fact that in conformal theories, the auxiliary field $G_{\mm}$
couples to matter exactly like a U(1) gauge field. To make this
obvious, it is convenient to modify the (2,0) supergravity covariant
derivative in \cite{H} by introducing a world-sheet U(1) generator
${\Hat {\cal Y}}$ and redefine $\nabla_A \to \nabla_A +
\d_A {}^{\mm} G_{\mm} {\Hat {\cal Y}}$. The resulting (2,0) supergravity
covariant derivative satisfies,
$$ \eqalign{ {~~~~}
[ ~ \nabla_+ \, , \, \nabla_+ ~\} &=~ 0 ~~~~,~~~~ [ ~ \nabla_+ \, , \, 
{\Bar \nabla}_+ ~\} ~=~ i \nabla_{\pp} ~~~~, \cr
[ ~ \nabla_+ \, , \, \nabla_{\pp} ~\} &=~ 0 ~~~~,~~~~ [ ~ \nabla_+ \, , \, 
{\nabla}_{\mm} ~\} ~=~ i \, {\Bar \S}^+ \, (~ {\cal M}
~+~ i \,{\Hat {\cal Y}}  ~)  ~~~~, \cr
[ ~ \nabla_{\pp} \, , \, \nabla_{\mm} ~\} &=~ \S^+ \, \nabla_+
~+~ {\Bar \S}^+ \, {\Bar \nabla}_+ ~+~ {\cal R}{\cal M} ~+~ {\cal F} 
{\Hat {\cal Y}} ~~~,
} \eqno(2.1)  $$
where $ \S^+ , ~ {\cal R}$ and ${\cal F}$ are superfields such that
$$ \eqalign{ {~~~~~~~~}
\S^+ | &\equiv~ [~  {\cal D}_{\pp} \chi_{\mm} ~-~ {\cal D}_{\mm} \chi_{\pp}
~-~ c_{\pp \, \mm} {}^c \chi_c {}^+ ~] ~~~~, \cr
{\cal R} | &\equiv~  r (e, \, \o(e, \chi )) ~+~ i \,[~ 
\chi_{\pp}{}^+ ({\Bar \S}^+ | ) ~+~ {\Bar \chi}_{\pp}{}^+ 
({ \S}^+ | )  ~] ~~~~, \cr
{\cal F} | &\equiv~ {\cal D}_{\pp} A_{\mm} ~-~ {\cal D}_{\mm} A_{\pp}
~-~ c_{\pp \, \mm} {}^c A_c  ~-~ [~ \chi_{\pp}{}^+ ({\Bar \S}^+ | )
~-~ {\Bar \chi}_{\pp}{}^+  ({ \S}^+ | ) ~ ] ~~~~,
} \eqno(2.2)  $$
and $r(e, \, \o(e, \chi ))$ denotes the world sheet curvature. The component
field content is just $(e_a {}^m , \, \chi_a {}^+ , \, A_a )$. The field
strength superfields satisfy
$$  {\Bar \nabla}_+ { \S}^+ ~=~ 0 ~~~~,~~~~ \nabla_+ { \S}^+ ~=~
{\cal R} ~+~ i\, {\cal F} ~~~~.
\eqno(2.3)  $$
The final result that is required to derive component results from (2,0)
superspace results is to note the ``density projection'' formulae;
$$ \eqalign{
\int d^2 \s d^2 \z^{\pp} ~ E^{-1} \, {\cal L} ~\equiv~ ~~
&\frac 12  \int d^2 \s d \z^+ ~ {\cal E}^{-1} \, (( \nabla_+ 
~-~ i \, 2 {\Bar \chi}_{\pp} {}^+ ) {\Bar \nabla}_+ {\cal L} ) |  \cr
+&\frac 12  \int d^2 \s d {\Bar \z}^+ ~ {\Bar {\cal E}}^{-1} \, (( 
{\Bar \nabla}_+ ~-~ i \, 2  \chi_{\pp} {}^+ ) {\nabla} _+{\cal L} ) | 
} \eqno(2.4) $$
valid for any (2,0) superfield ${\cal L}$ and as well
$$
\int d^2 \s d \z^+ ~ {\cal E}^{-1} \, {\cal L}_{- c} ~=~
\int d^2 \s ~ e^{-1} \, (( \nabla_+ 
~-~ i \, 2 {\Bar \chi}_{\pp} {}^+ ) \, {\cal L}_{- c} ) |
~~~~, 
\eqno(2.5) $$
valid for any chiral superfield $ {\cal L}_{- c}$.

One other interesting feature is the form of the (2,0) world sheet scale 
(Howe-Tucker) transformation laws of the covariant derivative. These take 
the forms
$$ \eqalign{
 \d_L \nabla_+ &=~ \frac 12 L \,  \nabla_+ ~-~ ( \nabla_+ f) {\cal M}
~-~ ( \nabla_+ g) {\Hat {\cal Y}} ~~~~, \cr
\d_L \nabla_{\pp} &=~ \frac 12 [ \, L~+~ {\Bar L} \,] \,  \nabla_{\pp} ~-~
i\, \frac 12 [\, {\Bar \nabla}_+ ( {L} ~-~ {\Bar f} ~-~ i {\Bar g} )]
\nabla_+  \cr
&{~~~~} ~-~ i \,
\frac 12 [\, { \nabla}_+ ( {\Bar L} ~-~ {f} ~+~ i { g} )] {\Bar \nabla}_+ \cr
&{~~~~} ~+~ i\,  [~ {\Bar \nabla}_+ \nabla_+ f ~+~  \nabla_+  {\Bar \nabla}_+
 {\Bar f}  ~]\,{\cal M} \cr
&{~~~~} ~+~ i\,  [~ {\Bar \nabla}_+ \nabla_+ g ~+~  \nabla_+  {\Bar \nabla}_+
 {\Bar g}  ~]\, {\Hat {\cal Y}} ~~~~, \cr
\d_L \nabla_{\mm} &=~ \frac 12 [ \, L~+~ {\Bar L} \,]
 \,  \nabla_{\mm} ~-~ ( \nabla_{\mm}
 F) {\cal M} ~-~ ( \nabla_{\mm} G) {\Hat {\cal Y}} ~~~,
} \eqno(2.6) $$
where the parameter superfields $L$, $f$, $g$, $F$ and $G$ are all expressed
in terms of a chiral superfield $\L$
$$ L \equiv~  \frac 12 (\, \L ~+~{\Bar \L}\,)  ~~~,~~~ f \equiv~ -
 \frac 12 \L  ~~~, 
~~~ F \equiv~ \frac 12 (\, \L \,+ \, {\Bar \L} \,) ~~~, $$
$$g \equiv~ -~ i \, \frac 12   ( \, 2 \L ~+~ {\Bar \L} \,)
~~~,~~~ G \equiv~ 0  ~~~. \eqno(2.7) $$
These imply the following transformations of the field strength $\S^+$
$$
\d_L \, \S^+ ~=~ \frac 32 \, L \, \S^+ ~+~ i
 ( \nabla_{\mm} \nabla_+ \L \,) ,
\eqno(2.8) $$
and we note that the transformation of the other two field strength superfields
follows from applying $\d_L$ to the latter result in (2.3).

In closing this section, we wish to return to the issue of uniqueness of
(2,0) supergravity. As we pointed out in the introduction to this
section, our first description of (2,0) supergravity did not include
a gauged U(1). Since from the view of string theory this U(1) appears
significant, it is certainly reasonable to ask if there are any
alternatives? The reason for asking this question is that the form
of the (2,0) supergeometry is ultimately responsible for the type of
space-time conformal compensator (see the discussion below) that is 
utilized. This question of uniqueness is also related to the question 
of whether there are more (2,0) scalar multiplets in addition to the 
chiral one?  The answer to both of these questions is yes. There is 
another way to take our initial construction of (2,0) supergravity 
and gauge U(1).  All that needs to be done is to take the covariant 
derivative in \cite{H} and change it according to $\nabla_A \to 
\nabla_A + \G_A {\Hat {\cal Y}}$ introducing a (2,0) matter vector multiplet 
($\G_A $) that is independent of (2,0) supergravity. The most important 
feature of this alternative description is that this theory possess a U(1) 
covariant superfield $G_{\mm}$ that is the lowest component of the 
supergravity field strength multiplet. Under this circumstance it can be 
shown that the Fradkin-Tseytlin term is given by 
$$
{\cal S}_{FT} ~=~ \int d^2 \s d^2 \z^{\pp} ~ E^{-1} \, \Phi \,
G_{\mm} ~~~, \eqno(2.9)
$$ 
where $\Phi$ is an arbitrary function of world-sheet scalar superfields.
This observation might open the way to alternative formulations of the 4D, 
N = 1 supergravity theory (non-minimal, new minimal) {\underline {derived}} 
from heterotic superstrings.

With regard to the existence of more (2,0) scalar multiplets, we have 
an answer that is derivable from some quite recent work on WZNW terms 
\cite{H1}. There it was shown that there exist a (2,2) scalar multiplet 
called the non-minimal scalar multiplet. This representation is distinct 
from chiral multiplets and possesses a (2,0) truncation and is thus a 
candidate to appear in a different world sheet action. The superfield 
description of this (2,0) multiplet requires two superfields, denoted 
by ${\rm Y}$ and $P_{\mm}$, that satisfy the constraint $ {\Bar \nabla
}_+ \nabla_{\mm} {\rm Y} = {\Bar \nabla}_+ P_{\mm}$. The free action 
for the multiplet is just
$$
\int d^2 \s d^2 \z^{\pp} ~ E^{-1} [\,  i {\rm Y} {\Bar P}_{\mm} ~+~ 
{\rm {h.\, c.}} \, ] ~~~~. \eqno(2.10)
$$

%%%%%%%%%%%%%%%%%%%%%%%%%%%%%%%%%%%%%%%%%%%%%%%%%%%%%%%%%%%%%%%%%%%%
\section{Manifest (SUSY$)^2$ vs. the New Green-Schwarz \newline
Action}
%%%%%%%%%%%%%%%%%%%%%%%%%%%%%%%%%%%%%%%%%%%%%%%%%%%%%%%%%%%%%%%%%%%

~~~~The (SUSY$)^2$ principle implies that the object of primary interest
is ${\cal Z}^{\un M}$ that maps from (2,0) superspace into 4D, N = 1
superspace. One choice to represent this map is ${\cal Z}^{\un M}
= ({\cal Z}^{\m} , \, {\cal Z}^{\dot \m} , \, {\cal Z}^{\m \dot \m})$
where ${\cal Z}^{\un M}$ is a (2,0) superfield, i.e. ${\Bar \nabla}_+
{\cal Z}^{\un M} = 0$. Since necessarily ${\cal Z}^{\un M}$ is complex
we may write
$${\cal Z}^{\m \dot \m} ~\equiv~ X^{\m \dot \m} ~+~ i\, Y^{\m \dot \m}
~~~~,~~~~ {\cal Z}^{\m } ~\equiv~ \frac 12 [\, \Q^{\m } ~+~ \D^{\m} \,]
~~~~,~~~~ {\cal Z}^{\dot \m } ~\equiv~ \frac 12 [\, {\Bar \Q}^{\dot \m } 
~-~ {\Bar \D}^{\dot \m} \,] ~~~~,
\eqno(3.1) $$
where $X^{\m \dot \m}$ and $Y^{\m \dot \m}$ are real.  We identify the 
usual 4D space-time superstring coordinates by
$$
\frac 12 [ {\cal Z}^{\m \dot \m} ~+~ ({\cal Z}^{\m \dot \m})^* ]
(\z^+ , \, {\Bar \z}^+ ,\, \s , \, \t) | 
 ~=~ \left(\begin{array}{cc}
~X^0 ~+~ X^3 & ~~X^1 ~-~ i X^2\\
{}~&~\\
~X^1 ~+~ i X^2 & ~~X^0 ~-~ X^3 \\
\end{array}\right)  {~~~~~~.}  
\eqno(3.2) $$
$$ [ {\cal Z}^{\m } \,+\,  ({{\cal Z}}^{\dot 
\m })^* ] (\z^+ , \, {\Bar \z}^+ ,\, \s , \, \t) |  ~=~ \Q^{\m} 
( \s , \, \t) ~~~.
\eqno(3.3) $$

As readily seen, manifest (2,0) supersymmetry has forced us to introduce
a sort of mirror superspace with coordinates $(\D^{\m} , \, Y^{\m 
\dot \m})$. We can turn this to our advantage. It is well known that
there are three bases in which to represent Salam-Strathdee superspace;
(a.) vector basis, (b.) chiral basis and (c.) anti-chiral basis.
By placing constraints on the mirror supercoordinates we seem able to
represent each of these. For example, the vector basis seems related
to the choice $ Y^{\m \dot \m} = {\D}^{\m } = 0$ and the
chiral basis seems related to $ Y^{\m \dot \m} = \Q^{\m } {\Bar \Q}^{\dot
\m} $ and ${\cal Z}^{\dot \m } = 0$ (note that ${\cal Z}^{\dot \m }$ is
not the conjugate of $({\cal Z}^{\m } )^*$). The chiral basis provides 
a minimal way in which to describe the superspace.

As the 2D world-sheet supergeometry is characterized by the (2,0)
supergravity covariant, $\nabla_A$, in order to describe the 4D, N = 1
space-time supergeometry, we introduce a vielbein $E_{\un M}{}^{\un A}$
that is a function of the coordinates ${\cal Z}^{\un M}$ and  
${\Bar {\cal Z}}^{\un M}$.  The quantities $\P_A {}^{\un A} \equiv
(\P_+ {}^{\un A} , \, \P_{\pp} {}^{\un A} , \,\P_{\mm} {}^{\un A}  \,)$
denote space-time supercovariant ``normals'' that are defined by,
$$\P_+ {}^{\un A} ~=~ (\, \nabla_+  {\cal Z}^{\un M} ) E_{\un M}{}^{\un A}
~~~,~~~
\P_{\pp} {}^{\un A} ~=~ (\, \nabla_{\pp}  {\cal Z}^{\un M} ) E_{\un M}
{}^{\un A} ~~~,~~~
\P_{\mm} {}^{\un A} ~=~ (\, \nabla_{\mm}  {\cal Z}^{\un M} ) E_{\un M}
{}^{\un A} ~~~,
\eqno(3.4) $$
and satisfy
$$
F_{A \, B}{}^{\un C} \equiv \nabla_A \P_B {}^{\un C} - (-)^{AB}
\nabla_B \P_A {}^{\un C} - T_{A \, B}{}^C \P_C {}^{\un C} - (-)^{A \un B}
\P_A {}^{\un A} \P_B {}^{\un B} T_{\un A \, \un B}{}^{\un C} = 0.
\eqno(3.5) $$
In other words, if $\P_{A}{}^{\un A} $ is regarded as a world-sheet 
gauge field, it has a vanishing field strength where $ T_{\un A \, 
\un B}{}^{\un C}$ acts as the structure constants for the gauge group. 
Alternately, we may regard $\P_{A}{}^{\un A} $ as a linear mapping 
operator that relates vectors and covectors on the superworld-sheet 
to those over the 4D, N = 1 super space-time via $e_A = \P_{A}{}^{\un A} 
E_{\un A}$ and $d \o^{\un A} = d \o^A \P_{A}{}^{\un A} $.  Although 
(3.5) is a classical equation, it is interesting to conjecture that its 
expectation value in a quantized theory is related to anomalies and 
critical dimensions.

Now having completed our definitions, we note that the remaining
component fields in ${\cal Z}^{\un M}$ (complex bosonic twistor
fields and complex NSR fermions) may be defined covariantly with
respect to both world-sheet and space-time manifolds through the
equation
$$ \P_{+}{}^{\un A} | ~ \equiv ~ (S_+ {}^{\a} , \, {\Tilde S}_+ 
{}^{\dot \a} , \, \psi_+ {}^{\un a} ) ~~~~.
\eqno(3.6) $$

For an action to describe the dynamics of ${\cal Z}^{\un M}$, we take
our motivation from the symmetries (both classical and quantum) of the 
action of reference \cite{F} and write
$$ \eqalign{
{\cal S} ~=~ &\Big\{ \int d^2 \s \, d^2 \z^{\pp} ~ E^{-1} \, [\,
{\Bar Z}^{\un M} E_{\un M}{}^{\un A} \P_{\mm}{}^{\un B} t^{(0)}_{\un A
\, \un B}   ~+~ \P_+ {}^{\un A} \P_+ {}^{\un B} \L_{\mm \mm
\un A \, \un B} \,] ~+~ {\rm {h. \, c.}}  \Big\} \cr
 +&\Big\{ \int d^2 \s \, d^2 \z^{\pp} ~ E^{-1} \, [\, \P_+ {}^{\un A}
\L_{\mm -}{}^{\un B} t^{(1)}_{\un A \, \un B} ~+~ 
\P_{\pp}{}^{\un A} \L_{\mm \mm}{}^{\un B} t^{(2)}_{\un A \, \un B} ~+~
\P_{\mm}{}^{\un A} \L_{}^{\un B} t^{(3)}_{\un A \, \un B} \,] ~+~
{\rm {h. \, c.}}  \Big\} \cr
+&\Big\{ \int d^2 \s \, d \z^+ ~ {\cal E}^{-1} \, [\, \S^+ \Phi({\cal 
Z}) \,]   ~+~ {\rm {h. \, c.}}  \Big\} ~~~~,
} \eqno(3.7) $$
where the quantities $t^{(i)}_{\un A \, \un B}$ are a set of constant
tensors. One parametrization of these is
$$
t^{(i)}_{\un A \, \un B} ~=~ 
\left(\begin{array}{ccc}
~k_1^{(i)} C_{\a \b} & ~~0 &  ~~0\\
~0 & ~~~k_2^{(i)} C_{\dot \a \dot \b} &  ~~0\\
~0 & ~~0 &  ~~ ~k_3^{(i)} C_{\a \b} C_{\dot \a \dot \b}\\
\end{array}\right)  {~~~~.~~~~~~~~~~~} 
\eqno(3.8) $$
For $k_1^{(0)} = k_2^{(0)} = 0$, $k_3^{(0)}$ nonvanishing, the first term 
produces the standard nonlinear $\s$-model with torsion for ${\cal Z}^{
\un M}|$ \cite{H,H2}.  For  $k_3^{(1)}$ nonvanishing, variation with respect 
to $\L_{\mm -}{}^{\un b}$ imposes the superfield equation $\P_+ {}^{\un b} 
= 0$. In the work of reference \cite{I}, the analog of this condition plays a 
critical role in eliminating would-be NSF fermions. Finally, for $k_1^{
(1)}$ and $k_2^{(3)}$ nonvanishing, a simple definition of propagators 
for the Grassmann coordinates ($\Q^{\a})$ is possible. We don't completely 
understand the role of $t^{(2)}_{\un A \, \un B}$, it seems related to the 
choice of basis (vector, chiral, anti-chiral).

No explicit factors of $\a'$ appear in our action. The reason for
this is that we can relegate all such factors to the zero modes
of ${\cal Z}^{\un M}$. By this we mean a mode expansion takes the
forms
$$ X^{\m \dot \m} ( \s , \, \t = \t_0 ) ~=~ ( 4 \p \a' )^{- \frac 12}
x^{\m \dot \m} ~+~ ... ~~,~~ \Q^{\m} ( \s , \, \t = \t_0 ) ~=~
( 4 \p \a' )^{- \frac 14} \q^{\m} ~+~ ... ~~.
\eqno(3.9) $$
where $...$ indicates higher mode terms. Using this convention,
$\a'$ never appears anywhere else in the formalism and all the two 
dimensional fields possess natural units of engineering dimensions
(i.e., 2d bosons = 0, 2d fermion = $\frac 12$).  Clearly, the
component level evaluation of the action is an important next
step. Since this promises to be quite intricate, we will carry
out this analysis in a future work. In closing this section, we
wish there to be no misunderstanding. We are {\underline {not}}
presently claiming that the action of (3.7) is the same as the 
$\s$-model of Berkovits and Siegel. Instead we propose it as
the starting point in trying to construct a classical manifestly
(SUSY$)^2$ model that matches many properties of their construction.

%%%%%%%%%%%%%%%%%%%%%%%%%%%%%%%%%%%%%%%%%%%%%%%%%%%%%%%%%%%%%%%%%%%%
\section{Will the Real 4D, N = 1 Supergravity Limit of \newline
Heterotic String Theory Please Stand up?}
%%%%%%%%%%%%%%%%%%%%%%%%%%%%%%%%%%%%%%%%%%%%%%%%%%%%%%%%%%%%%%%%%%%

~~~~There are many reasons why a complete manifest realization of
superstring theory is desirable.  Presently, many misunderstandings
exist due to such a complicated theory being formulated in
such an incomplete manner.  An example of this arose several
years ago \cite{J,K} regarding the pure 4D, N = 1 supergravity
limit of heterotic superstring theory.  At first \cite{J} it was 
argued that the ``new minimal'' off-shell version of 4D, N = 1 
supergravity must {\underline {necessarily}} occur at this limit. It 
was later \cite{K} noted that such a proposal was inconsistent  with 
the superspins implied by independent superstring theory arguments.
In \cite{G}, an analysis was performed to see how these two
competing claims could occur and, remarkably enough, it was shown
that there existed an ambiguity in the interpretation of the results 
of \cite{J}. Those results are equivalent to a derivation of the 
form of the graded commutator algebra of the superspace supergravity
covariant derivative. In \cite{G} it was shown that the superspace 
supergravity covariant derivative thus derived could be expressed as
either the ``new minimal'' off-shell version of 4D, N = 1 supergravity
or as the ``old minimal'' off-shell version of 4D, N = 1 supergravity
``entangled'' with a tensor multiplet that is also used as a composite
connection for ${\cal R}$-symmetry.  This last interpretation may seem 
counter intuitive and unnatural but it provided the only logical way
to reconcile the different claims \cite{J,K}. On the basis of 4D, N 
= 4 heterotic string theory, we provided a justification for why this
bizarre structure must arise. From all of our previous investigations 
of 4D, N = 4 supergeometry \cite{K1}, it can be seen that this composite 
U(1) connection was always present (either implicitly or explicitly). 
It was therefore natural to conclude that since 4D, N = 1 heterotic 
strings are closely related to 4D, N = 4 heterotic strings, a remnant 
of the N = 4 U(1) composite connection could occur in the N = 1 theory.

The new Green-Schwarz formulation has now completely 
{\underline {vindicated}} our deductive reasoning! It has now been
{\underline {rigorous}} {\underline {derived}} that the pure 4D, N 
= 1 supergravity limit of the heterotic string is (written in
{\it {Superspace}} conventions as in \cite{G})
\begin{center}
{\bf {4D, N = 1 $\b$FFC Supergeometry }} 
\end{center}
$$
\eqalign{
[ \nabla_{\a }, \nabla_{\b} \}  &= 0   ~~, ~~ \cr
[ \nabla_{\a }, { \bar {\nabla}}_{\dot \a} \}  &=~  i {\nabla}_{ \un a}
~+~ H_{\b \dot \a} {\cal M}_{\a } {}^{\b} ~-~ 
H_{\a \dot \b} {\bar {\cal M}}_{\dot \a } {}^{\dot \b}~~~,~~ \cr
[ \nabla_{\a }, \nabla_{\un b} \}  &=~
 i ( \nabla_{\b} H_{\g \dot \b}) {\cal M}_{
\a} {}^{ \g} ~-~ i \frac 12 
( \nabla_{  ( \a} H_{\b ) \dot \g}) {\bar {\cal M}}_{ \dot \b} 
{}^{ \dot \g}    ~+~ i \frac 12 
( \nabla_{  ( \a} H_{\b ) \dot \b}) {\cal Y}  ~~ \cr
&~~~~~+~ i C_{\a \b} [~ {\bar W}_{\dot \b \dot \g} {}^{\dot \d}
 { \bar {\cal M}}_{\dot \d} {}^{\dot \g} ~+~  \frac 16 
 \nabla^{\d} H_{\d \dot \g} {\bar {\cal M}}_{ \dot \b} 
{}^{ \dot \g} ~-~ \frac 12 \nabla^{\b} H_{\b \dot \b} {\cal Y} 
~] ~~~, {~~~~~~~\,~\,~}\cr } $$
$$\eqalign{ {~~~~}
[ { \nabla}_{\un a }, \nabla_{\un b} \}  ~=~ 
\big \{ &~ i \frac12  C_{\a \b } H^{\g} {}_{ ( \dot \a} \nabla_{ \g 
\dot \b)} ~-  \frac 14 C_{\a \b} [~ {\bar \nabla}_{ ( \dot \a} 
\nabla^{\d} H_{\d \dot \b ) } ~+~ i 2 \nabla^{\g} {}_{ ( \dot 
\a } H_{\g \dot \b) } ~ ] {\cal Y}\cr
&+~ [~ C_{\dot \a \dot \b} ( W_{ \a \b} {}^{ \g} ~-~ 
\frac 16 ( {\bar \nabla}^{\dot \g} H_{ ( \a \dot \g}) \d_{ \b )} {}^{\g}
) ~-~ \frac 12 C_{\a  \b} ( {\bar \nabla}_{ ( \dot \a } 
H^{\g} {}_{\dot \b ) })~] \nabla_{\g} \cr
&-~ [~ C_{\dot \a \dot \b} W_{ \a \b \g \d} ~+~ \frac 14 C_{\g (\a | }
({\nabla}_{ \d} {\bar \nabla}^{\dot \d } H_{ | \b ) \dot \d}) ~~\cr
&~~~~~~~~~+\frac 14 C_{\g (\a | }
({\bar \nabla}^{ \dot \d}
{\nabla}_{| \b ) } H_{ \d \dot \d})~] {\cal M}^{\g \d} \cr
&+\frac 16 C_{\g (\a }  C_{\b ) \d } [~
({\nabla}^{ \e}
{\bar \nabla}^{\dot \e } H_{ \e \dot \e})~] {\cal M}^{\g \d} \cr
&+ \frac 12 C_{\a \b } [~{\nabla}_{ \g} {\bar \nabla}_{ ( \dot \a } 
H_{\d} {}_{\dot \b ) } ~] {\cal M}^{\g \d}
~+~ h.c.~~~ \big \} ~~,  \cr }
\eqno(4.1)$$
$$ 
H_{\a \b \g} ~=~ H_{\a \b \dot \g} ~=~ H_{\a \b \g} ~=~ H_{\a \b 
\un c} ~~~,~~~ H_{\a \dot \b \un c} ~=~ i \frac 12 C_{\a \g} C_{\dot \b \dot \g}
~=~ 0 ~~~, $$
$$ 
{H}_{\a \un b \un c} ~=~ 0 ~~,~~ {H}_{\un a \un b \un c} 
~=~ i \frac 14  [~ C_{\b \g} C_{ \dot \a ( \dot \b } {H}_{\a \dot 
\g )} ~-~ C_{ \dot \b \dot \g } C_{ \a ( \b } {H}_{ \g ) \dot \a
} ~ ]  ~~. 
\eqno(4.2) $$

The superspace torsions $(T_{\un A \, \un B}{}^{\un C})$, Lorentz
curvatures $(R_{\un A \, \un B \, \g}{}^{\d}$ and $R_{\un A \, \un B 
\, \dot \g}{}^{\dot \d})$ and ${\cal R}$-symmetry field strength
$(F_{\un A \, \un B })$ can be read off by noting that in general
we have
$$
\Big[ \, \nabla_{\un A} \, , \, \nabla_{\un B} \, \Big\} ~=~
T_{\un A \, \un B}{}^{\un C}  \nabla_{\un C} ~+~
R_{\un A \, \un B \, \g}{}^{\d} {\cal M}_{\d} {}^{\g} ~+~
R_{\un A \, \un B \, \dot \g}{}^{\dot \d}{\cal M}_{\dot \d} {}^{\dot \g}
~+~ F_{\un A \, \un B } {\cal Y} ~~~, 
\eqno(4.3) $$
where ${\cal M}_{\d} {}^{\g}$ and ${\cal M}_{\dot \d} {}^{\dot \g} $
refer to the anti-self dual and self-dual parts, respectively, of the
of the Lorentz generators multiplied by Pauli matrices. In the same vein,
${\cal Y}$ refers to the generator of ${\cal R}$-symmetry. Equations
(4.1) and (4.2) show that all of the geometrical quantities are
expressed solely in terms of $H_{\un a}$ (at lowest order in $\q$
the axion field strength), $W_{\a \, \b \, \g}$ (at lowest order
in $\q$ the gravitino field strength) and their spinorial derivatives.
The most remarkable feature of (4.1) and (4.2) is the fact that they
do {\underline {not}} contain the auxiliary field multiplets ($G_{\un
 a}$ and $R$)!

Similarly, the general expression of the axion multiplet field strength
defined by $H_{\un A \, \un B \, \un C}$ whose components for various 
choices of Lorentz indices is explicitly given by,
$$ \eqalign{
H_{\a \, \b \, \g} &\equiv~ \frac 12 \nabla_{(\a |} B_{ |\b \g )}
~-~ \frac 12 T_{( \a \, \b |}{}^{\un E} B_{\un E | \g) } ~~~~, \cr
H_{\a \, \b \, \dot \g} &\equiv~  \nabla_{(\a |} B_{| \b ) \dot \g }
~+~ \nabla_{\dot \a } B_{\a  \b } ~-~ T_{\a \, \b}{}^{\un E} B_{\un E \dot \g }
~-~ T_{\dot \g ( \a |}{}^{\un E} B_{\un E | \b) } ~~~~, \cr
H_{\a \, \b \, \un c} &\equiv~  \nabla_{(\a |} B_{ | \b ) \un c }
~+~ \nabla_{\un c} B_{\a  \b } ~-~ T_{\a \, \b}{}^{\un E} B_{\un E \un c }
~-~ T_{\un c ( \a |}{}^{\un E} B_{\un E | \b) } ~~~~, \cr
H_{\a \, \dot \b \, \un c} &\equiv~  \nabla_{\a } B_{\dot \b  \un c } ~+~
\nabla_{\dot \b } B_{\a  \un c } ~+~ \nabla_{\un c} B_{\a  \dot \b }
~-~ T_{\a \, \b}{}^{\un E} B_{\un E \un c } \cr
&{~~~~~} - T_{\un c  \a }{}^{\un E} B_{\un E \dot \b }
~-~  T_{\un c \dot \b}{}^{\un E} B_{\un E  \a } ~~~~, \cr
H_{\a \, \un b \, \un c} &\equiv~  \nabla_{\a } B_{\un b  \un c }
~-~ \nabla_{[ \un b | } B_{\a | \un c ]} ~-~ T_{\a \, [ \un b |}{}^{\un E} 
B_{\un E | \un c] } ~-~  T_{\un b \un c}{}^{\un E} B_{\un E  \a } ~~~~, \cr
H_{\un a \, \un b \, \un c} &\equiv~ \frac 12 \nabla_{[ \un a |} B_{| \un b
 \un c ]} ~-~ \frac 12 T_{[ \un a \, \un b|}{}^{\un E} B_{\un E | \un c] } 
~~~~. \cr
} \eqno(4.4) $$
Here $B_{\un A \, \un B }$ refers to a super 2-form whose rigid
geometry was given in \cite{L} and whose local supergeometry, implied the 
4D, N = 1 heterotic string, can be read by comparing (4.2) with (4.4).
All components of the field strength not explicitly written above may
be obtained by complex conjugation.  

Although the constraints in (4.1) and (4.2) are the most convenient
from the view of heterotic string theory, they are by no means unique.
As shown in \cite{G}, there exist field re-definitions that we call
``entangling'' that can be used to relate these to any specific
supergeometry that contains minimal off-shell supergravity plus a 
tensor multiplet. Along the lines of a historic perspective, we note
that the first appearance of this class of superspace geometries
was within the context of 10D, N = 1 $\b$-function calculations
\cite{M}. The analogs of these constraints were found to have the
consequence that the entire one-loop contribution to the $\b$-function
comes from a single graph. For this reason, the constraints have been
called the $\b$FFC (beta function favored constraint) supergeometry.
The results of (4.1) and (4.2) are the direct descendants of their
10D progenitors and a major discovery of \cite{G} was to show that
the inclusion of the composite ${\cal R}$-symmetry connection allowed
these to appear in 4D, N = 1 supergeometries.

A few words are in order as to how the work of Berkovits and Siegel
rigorously leads to (4.1). As noted in \cite{G}, the standard 4D, N = 
1 GS $\s$-model action actually possesses {\it {spacetime}} superconformal
symmetry.   As such, there is no way to distinguish what set of auxiliary 
fields are associated with the 4D, N = 1 supergravity theory coupled to
the GS action.  The Berkovits-Siegel action explcitly breaks the
spacetime superconformal symmetry by coupling a scale compensator
superfield ({\underline {not}} the usual dilaton) in the Fradkin-Tseytlin 
term. It is a well known result \cite{L1} of supergravity theory that 
a given Poincar\' e supergravity theory is associated with a given 
choice of scale compensator. In particular, if the scale compensator
is chiral, the resulting Poincar\' e supergravity theory must be
the ``old minimal'' theory. The FT term in \cite{F} can 
{\underline {only}} accommodate a chiral compensator and thus 
the pure supergravity sector of the 4D, N = 1 heterotic string is 
the old minimal theory.

One final noteworthy consequence of the now rigorous derivation of the
pure 4D, N = 1 supergravity limit has to do with results that have
previously been accepted as facts about the phenomenological relevant
form of the low-energy effective action. In a large part of the literature
on string-inspired model building, the axion is represented as a 
pseudo-scalar that is part of a chiral superfield. The new results 
suggest that with the requirement of manifest 4D, N = 1 supersymmetry,
the axion must necessarily be represented as a 2-form whose field strength 
couples to matter as the gauge-field for ${\cal R}$-symmetry. At a 
minimum, issues that until now have been considered settled must be 
re-examined.

%%%%%%%%%%%%%%%%%%%%%%%%%%%%%%%%%%%%%%%%%%%%%%%%%%%%%%%%%%%%%%%%%%%%
\section{The $\k$-symmetry Transformation: Birth, Death \newline
and Resurrection}
%%%%%%%%%%%%%%%%%%%%%%%%%%%%%%%%%%%%%%%%%%%%%%%%%%%%%%%%%%%%%%%%%%%

~~~~One of the fascinating points regarding the new covariant formulation
of 4D, N = 1 superstrings is the fate of $\k$-symmetry.  This symmetry
originally was found in the superparticle action \cite{N} and was
later interpreted to be related to twistor transformations \cite{O}.
In a prior attempt to use Batalin-Vilkovisky quantization \cite{P},
it was found that $\k$-symmetry was the primary villain that prevented
a successful quantization \cite{Q}. On the otherhand, within our
prior investigation of 4D, N = 1 GS actions, $\k$-symmetry was an
important tool that was responsible for the correct deduction of the
form of the pure 4D, N = 1 supergravity limit of heterotic string theory. 
So it is useful to revisit this issue in light of the new formulation.

Foremost, we observe that there is {\underline {no}} invariance in the
complete 4D, N = 1 theory formulated by Berkovits, that corresponds
to $\k$-symmetry. Never the less, a part of the total action {
\underline{does}} realize $\k$-symmetry, i.e. a sector of the total
action is $\k$-symmetry invariant.  Some years ago, we suggested that
a completely consistent and manifestly supersymmetric formulation of
4D superstrings would possess this property \cite{R} and named those
terms of the complete action with this property, the ``kernel''
of the theory. Let us be a bit more explicit, if one writes out the
complete action of \cite{F}, then one finds it can be written
(with an appropriate change of notation) as the terms included in (5.2) 
below plus many other terms (e.g. the FT-term. etc.). Only (5.2) 
constitutes the kernel.

The most general kernel of a 4D GS $\s$-model can be constructed as
follows.  Introduce world-sheet fields (that must ultimately be embedded
into a larger theory) ${\cal Z}^{\un M} \equiv ( \Q^{\m \, i}, \,
\Q^{\m \, i'} , \, {\Bar \Q}^{\dot \m}{}_{ i}, \, {\Bar \Q}^{\dot \m}
{}_{ i'} , \, X^{\m \dot \m})$. The Grassmann coordinates $(\Q^{\m \, 
i}(\s, \t)$ and $\Q^{\m \, i'}(\s, \t)$) are introduced in the form of 
two-component spinors which carry additional ``isospin'' indices $i$ 
and $i'$ (where $i = 1, ..., N_L$ and $i' = 1, ..., N_R$ for some 
integers $N_L$ and $N_R$). We introduce the symbol ${\Hat H}_{\un A \, 
\un B \, \un C}$ defined by
$$
{\Hat H}_{\un A \, \un B \, \un C} ~=~
i \frac 12 C_{\a \g} C_{\dot \b \dot \g}  \left\{ \matrix{
~
{\d}_{i} {}^{j} ~~ &~~: if~ {\un A} 
= \a~ i ~,~ {\un B} = {\dot \b}~ j ~,~ {\un C} = \g \dot \g ~  \cr 
       ~ $~~$ ~~ &~~ {\rm or~any~even~permutation,} ~~~~~~~~~ \cr
~ -  {\d}_{i}{}^{j}
 ~~ &~~: {\rm for~any~odd~permutation,} ~~~~~~~ \cr
~ -  {\d}_{i'} {}^{j'} ~~ &~~: if~ {\un A} 
= \a~ i' ~,~ {\un B} = {\dot \b}~ j' ~,~ {\un C} = \g \dot \g ~ \cr 
       ~ $~~$ ~~ &~~ {\rm or~any~even~permutation,} ~~~~~~~~~ \cr
~   {\d}_{i' }{}^{j'}
 ~~ &~~: {\rm for~any~odd~permutation,} ~~~~~~~ \cr
0 ~~ &~~:~~{\rm otherwise.}   ~~~~~~~~~~~~~~~~~~~~~~~~~
\cr } \right\}    ~~. 
\eqno(5.1) $$
and using this symbol the kernel takes a universal form
$$
{\cal S}_{GS}^{Kernel} ~=~ \int d^2 \s ~ e^{-1} \, \Big[ ~ - \P_{\pp}
{}^{\un a} \P_{\mm}{}^{\un b} \eta_{\un a \un b} ~+~ \int_0^1 d y {\Hat 
\P}_y {}^{\un A} \P_{\pp}{}^{\un B}\P_{\mm}{}^{\un C} {\Hat H}_{\un A 
\, \un B \, \un C} ~ \Big] ~~~~,$$
$$ {\Hat {\cal Z}}^{\un M} ~\equiv~ {{\cal Z}}^{\un M} (\s, \, \t , \,
y) ~~~~,~~~~ \P_y {}^{\un A} ~\equiv~ (\pa_y \, {{\cal Z}}^{\un M} \,)
E_{\un M} {}^{\un A} ({\Hat {\cal Z}}) ~~~~,~~~~ {\Hat H}_{\un A \, 
\un B \, \un C} ~\equiv~ {\Hat H}_{\un A \, \un B \, \un C}({\Hat {\cal 
Z}}) ~~~~.
\eqno(5.2) $$ 
where we have used the Vainberg construction to express the action in
terms of the field strength of the 2-form \cite{S}.

As shown in \cite{Q}, for arbitrary values of $N_L$ and $N_R$ it is possible
to define $\k$-symmetry variations that leave the kernel invariant. The
reason this is interesting is because the number of space-time
supersymmetries, N is the sum $N_L + N_R$. We pick $N_L = 1, \,
N_R = 0$ to define the 4D, N = 1 theory. For N = 2, there are
two choices, either $N_L = 2, \, N_R = 0$ or $N_L = 1, \, N_R = 1$.
Precisely, these cases can be seen in the recent work of 
\cite{F}. Apparently, the first of the two choices corresponds to
the heterotic compactification and the second to a type-II compactification.
The fact that this remnant of structures found through the use of
$\k$-symmetry arguments survives into the full theory provides us
with a second example showing that though $\k$-symmetry is broken,
its use can still lead to useful insights and questions.  Of course
we expect these results to generalize to higher values of $N_L$ and
$N_R$.  For example, for N = 3, there are two different $(N_L , \, N_R)$
possibilities; (3,0) and (2,1). The first is clearly a heterotic
compactification. The second, however, is quite mysterious. All
type-II compactifications are expected to have $N_L = N_R = \frac 12 N$.
Considering the case of N = 4 there are possibilities; (4,0), (3,1)
and (2,2). Finally for N = 8, we see the possibilities, (8,0), (7,1),
(6,2), (5,3) and (4,4). Once again we identify the first with the
heterotic case and the last with the type-II case. The others are
again mysterious. Should the intermediate cases prove to lead to
consistent superstrings, they would be examples of 4D superstrings
that do not have their origins in 10D.

%%%%%%%%%%%%%%%%%%%%%%%%%%%%%%%%%%%%%%%%%%%%%%%%%%%%%%%%%%%%%%%%%%%%
\section{Solving the 4D, N = 1 Conformal Symmetry \newline
Problem and Beyond}
%%%%%%%%%%%%%%%%%%%%%%%%%%%%%%%%%%%%%%%%%%%%%%%%%%%%%%%%%%%%%%%%%%%

~~~~In a previous work \cite{G}, we found a puzzling situation. For the
usual 4D, N = 1 GS non-linear $\s$ model, the condition of $\k$-symmetry
invariance implied that the 4D, N = 1 superspace could describe any
space-time superconformal background. At the time of our discovery of
this fact, we pointed out that this situation clearly must be resolved
in order to have a 4D, N = 1 heterotic string whose point particle limit
yielded Poincar\' e as opposed to conformal supergravity theory.

Once again, the work by Berkovits and Siegel provides a simple resolution
to this problem. By introducing the covariant scale compensator as the
coefficient of the world sheet curvature in the Fradkin-Tseytlin term, 
the apparent superconformal symmetry of the 4D, N = 1 GS non-linear
$\s$-model is easily broken in precisely the same manner that conformal
symmetry is broken in superfield supergravity theories. This suggests
that 4D, N = 1 superstring theory (and quite likely all superstring
theories) follow the paradigm of the superfield supergravity formulation
involving pre-potentials. We had long conjectured that this would
be the case \cite{D}.

A final remaining challenge we must undertake is in the realm of
compactification.  Although the work of Berkovits and Siegel only
contains explicit results for a Calabi-Yau compactified sector, 
there are good reasons to derive {\underline {explicit}} results for
{\underline {other}} types of compactifications (some of which cannot 
even be interpreted as higher D theories). Phenomenologically, there 
is no guarantee that the simplest string-extended standard model 
is a member of the class of Calabi-Yau compactifications.  As we have 
noted in our previous $\s$-model investigation \cite{T} of Calabi-Yau 
compactification, CY $\s$-models seem to be only a special case of 
a more general class of models.  In fact, it appears that our work
within the NSR formulation of the $\s$-model has the interesting
feature that its Calabi-Yau sector is exactly the same as that
of the new GS formulation \cite{U}. We therefore believe that
this property will be true of any consistently formulated NSR
compactification sector!  The most general members of the $\s$-model
compactification sectors we have found are the Lefton-Righton Thirring 
Models (LRTM) \cite{D,V}. So in a future work we will investigate the 
implication of the new GS formulation within the (2,0) LRTM class.

\noindent
%%%%%%%%%%%%%%%%%%%%%%%%%%%%%%%%%%%%%%%%%%%%%%%%%%%%%%%%%%%%%
{\bf {Acknowledgment; }} \newline \noindent The author wishes
to thank N. Berkovits and W. Siegel for enlightening conversations
about the new manifestly supersymmetric string theory formulation.
%%%%%%%%%%%%%%%%%%%%%%%%%%%%%%%%%%%%%%%%%%%%%%%%%%%%%%%%%%%%%

\newpage
%%%%%%%%%%%%%%%%%%%%%%%%%%%%%%%%%%%%%%%%%%%%%%%%%%%%%%%%%%%%%%%%%%%%
%\sect{References}
%%%%%%%%%%%%%%%%%%%%%%%%%%%%%%%%%%%%%%%%%%%%%%%%%%%%%%%%%%%%%%%%%%%%

\end{document}